\documentclass{ws-procs10x7}
\usepackage{balance}

\newcolumntype{d}[1]{D{.}{.}{#1}}

\def\Journal#1#2#3#4{{\it #1} {\bf #2}, #3 (#4)}

\makeindex
\begin{document}

\title{FORWARD PROTON DETECTORS AT HIGH LUMINOSITY AT THE LHC}

\author{M. GROTHE}

\address{Dipartimento di Fisica Sperimentale, Universit\`a  di Torino, via P. Giuria 1, 10125 Torino, Italy\\
E-mail: Monika.Grothe@cern.ch}

 
\twocolumn[\maketitle\abstract{We discuss the special challenges posed by measuring 
diffractive and forward physics at the LHC at high luminosity and the solutions 
proposed by the FP420 R\&D collaboration.}
]

\section{Introduction}

The term ``high luminosity'' throughout this article is taken to mean
``luminosities at which event pile-up is significant''. In that case,
the rapidity gaps typical of diffractive events tend to be filled in by
particles from the overlaid pile-up events. Hence the selection of diffractive
events has to rely entirely on tagging the diffractively scattered protons
which escape intact down the beam-pipe.

In recent years, theoretical advances have identified forward proton tagging 
at the 
LHC as a promising tool for the search for and the characterization
of new particles at the LHC.\cite{KMR} 
At the highest LHC luminosities,
central exclusive production (CEP) 
may become a tool for the discovery of any object that 
couples to gluons. A case in point is a Standard Model or MSSM Higgs boson 
with mass close to the current
exclusion limit. CEP would provide a clean 
experimental signature for its discovery; would make possible an unambiguous 
determination of its quantum numbers; and would offer, as the sole channel at 
the LHC, a direct means of measuring the CP structure of the Higgs 
sector.\cite{Forshaw}

Forward proton tagging at high luminosities will also give access to a rich 
QCD program. Measuring diffraction in the presence of a hard scale
will render possible the investigation of fundamental aspects
of the proton structure as pioneered at HERA and the TEVATRON. Diffractive 
Parton Distribution Functions provide a view on the proton
through a lens that filters out everything but the vacuum quantum numbers;
Generalized Parton Distributions contain information on the correlations
between partons in the proton. Along with these, the so-called rapidity gap
survival probability can be experimentally determined. This quantity is 
closely linked to soft rescattering effects and to the features of the 
underlying event at the LHC.\cite{diffprogram}

\section{The FP420 R\&D project}

For slightly off-momentum protons, the LHC beamline with its magnets is
essentially a spectrometer. If diffractively scattered protons are bent 
sufficiently, but little enough to remain within the beam-pipe,
they can be detected by means of detectors inserted into the beam-pipe and 
approaching the beam envelope as closely as possible.

At nominal LHC optics, an ideal position is at a distance of $\pm420$~m from the
interactions points of the ATLAS and CMS experiments, where coverage in the 
fractional momentum loss $\xi$ of the proton of $0.002 < \xi < 0.02$ can be 
achieved.

The FP420 R\&D collaboration, with members from ATLAS, CMS and the LHC, aims at
installing high precision tracking and fast timing detectors close to the beam at
the 420~m location.\cite{FP420LOI} The currently considered configuration foresees, over a length
of 10~m, 2 or 3 stations of Silicon detectors with about 8 layers each.
Fast timing detectors complement this setup. 
They would be capable of determining, within a resolution of a few millimeters,
whether the tagged proton candidates came from the same vertex as the hard scatter.

The region at $\pm 420$~m from the IP of ATLAS or CMS is located in the cold section
of the LHC. The engineering challenge of integrating detectors operated at room 
temperature into a beamline operated at cryogenic temperatures can be resolved
by effecting a cold-warm transition by means of modifying an already existing
LHC beamline element, the LHC Arc Termination 
Module. The engineering design work is already well underway.

The currently preferred solution for a mechanism to insert the tracking detectors
into the beam-pipe and to approach them to the beam to within about 3~mm is a 
movable beam-pipe section to which the detector stations would be attached. 
Beam pipe section and detectors would be kept in a position remote from the 
beam during injection, and would be moved as close as possible to
the beam-line once the beam has stabilized and the narrow beam envelope of
collision running has been reached.

The tracking detectors in the 420~m location have to be sufficiently radiation
hard to be operable in the immediate vicinity of the LHC beam, and their
insensitive area on the side closest to the beam envelope needs to be as small as
possible to maximize acceptance. The technology currently foreseen is edgeless
3-D Silicon where the electrodes are processed inside of the Silicon bulk
instead of being implanted on its surface and where the width of the insensitive
volume at the edge is smaller than $5~\mu$m. In addition, this novel type of 
Silicon detectors has been found to withstand the dose expected 
at 420~m for an integral of 5 years of LHC running at 
10 times the LHC design luminosity. The 
current prototypes utilize the radiation-hard ATLAS pixel readout chip and were 
tested in a test beam at CERN this summer. 

With a Silicon detector electrode pitch of $50~\mu$m, a resolution in the two 
spatial 
dimensions of about $15~\mu$m can be reached. Preliminary Monte-Carlo simulations 
indicate that for CEP of a Higgs boson with mass 
between 120~GeV and 200~GeV, this translates into a mass resolution of around 
1.5~GeV, when the two protons are both detected at 420~m (420+420 tag).

\section{Pile-up background}

Single diffractive events contribute about 15\% of the inclusive QCD cross section 
at the LHC, double Pomeron exchange events a few percent. In addition, a certain
portion of non-diffractive events contain leading protons with a fractional
momentum loss small enough that they can be observed in near-beam detectors.
These generally soft events with leading protons are present in the event pile-up.

The unprecedented high luminosities at the LHC come at the cost of event pile-up,
i.e. each hard scatter will be overlaid with a luminosity-dependent number of 
generally soft events. At an instantaneous luminosity of 
$2 \times 10 ^{33} {\rm cm}^{-2}{\rm s}^{-1}$, the average number is 7 events per
crossing, at $1 \times 10 ^{34} {\rm cm}^{-2}{\rm s}^{-1}$ it is 35.
Of these pile-up events, of the order 1\% contain a proton within the acceptance of
near-beam detectors at 420~m from the IP. 

For the selection of diffractive events at the LHC at high luminosities, these
leading protons from pile-up are a major background source. To illustrate the 
point, we consider the case of a Higgs boson with 120~GeV mass that decays 
into a pair of $b$-jets. For the non-diffractive production of a Higgs boson of this
mass and decay channel the background from inclusive dijet production is 
overwhelming. Quantum number selection rules in the case of CEP 
(i.e. double Pomeron exchange) suppress this background to a large
extent. However, inclusive dijet events, when they occur in coincidence with 
pile-up events that have leading protons within the acceptance of the near-beam 
detectors at $\pm 420$~m, again appear to have the same signature as the signal. 
Simple combinatorics yields as estimate at 
$2 \times 10 ^{33} {\rm cm}^{-2}{\rm s}^{-1}$ that of the order of a few per mill
of inclusive dijet events are being mistaken as signal events.
Given the much larger cross section of inclusive dijet events compared to the 
signal, this is the most important source of background.

This background can be reduced by exploiting the 
correlations between quantities measured in the main detector and those measured
with the near-beam detectors. One possibility is to estimate the fractional 
momentum loss, $\xi$, of the protons with the help of the dijet system as
$\xi_{1,2} = \frac{1}{\sqrt{s}} [\Sigma E_T^{jet} e^{\pm \eta}]$, where the sum is over
the two jets and $\eta$ denotes their pseudorapidity.

Another possibility is the use of fast timing detectors that determine whether the
protons seen in the near-beam detectors came from the same vertex as the hard
scatter. Fast timing detectors with an expected vertex resolution of better than 
3~mm are part of the FP420 project. Preliminary Monte-Carlo studies 
indicate that with nominal LHC running conditions a rejection of about 97\% is 
possible 
of events that appear to be double Pomeron exchange events, but where the
protons in reality originated from coincidences with pile-up events.

Two protoypes are currently under discussion, one using Quartz as Cherenkov
medium, the other gas. Both utilize micro channel plate
photo-multipliers which are known to have yielded a time resolution of about 10~ps in Cherenkov-light based Time-of-Flight detectors. Results from test-beam 
measurements at Fermilab with the two prototypes this summer will be available 
soon.

\section{Detectors at 220~m distance}

Additional near-beam detectors closer to the IP than 420~m would increase the 
physics reach for forward and diffractive physics, notably with respect to 
triggering and with respect to the acceptance for 
centrally produced mass states of higher values.
A suitable location is at 
$\pm 220$~m from the ATLAS or CMS IP. At nominal LHC optics, detectors there
provide an acceptance of 
$0.02 < \xi < 0.2$, i.e. quite complementary to the acceptance at $\pm 420$~m.

In CEP, the mass of the centrally produced
system, $M$, and the fractional momentum loss, $\xi_1, \xi_2$, of the two protons are 
correlated via: $M^2 = \xi_1 \xi_2 s$, where $\sqrt{s}=14$~TeV at the LHC.
This translates into a minimum observable mass of about 30~GeV, when
the protons are seen in the 420~m detectors on each side. For masses
above about 80~GeV, a marked increase in the efficiency 
results from events with asymmetric $\xi$ values, where
one proton is observed at 420~m and the other at 220~m. At $M=200$~GeV, for 
example, 420+420 tags limit the selection efficiency to about 10\%, while it
is almost 90\% for 420+220 tags.\cite{Boonekamp}

Preliminary Monte-Carlo studies indicate for CEP of a Higgs boson that the 
achievable mass resolution
with 420+220 tag varies between less than 3~GeV (3.5~GeV) for ATLAS (CMS) 
at a Higgs mass of 120~GeV, and less than 2.5~GeV (3~GeV) for ATLAS (CMS) at a 
Higgs mass of 200~GeV.

The 420~m location is too far away for signals to be processed within the first 
level trigger (L1) latency at both ATLAS and CMS. Hence, events with protons tagged
at 420~m need to be triggered either with the central detector alone or with 
near-beam detectors closer to the IP. 

In the case of CEP of a Higgs boson of 120~GeV mass 
decaying into two $b$-jets,
triggering with central detector conditions alone is only possible with high $p_T$
muons. About 10\% of signal events are accepted by the L1 muon trigger stream.
The L1 thresholds foreseen for dijets are generally too high.
Adding a L1 trigger stream that combines a dijet condition with requiring a tag
on one side at 220~m makes it possible to lower the dijet L1 thresholds 
sufficiently to retain about 10\% of the signal events, while the output rate 
stays within the L1 bandwidth limits for luminosities up to 
$2 \times 10^{33} {\rm cm}^{-2} s^{-1}$.\cite{trigger}

The option of placing near-beam detectors at $\pm 220$~m from the ATLAS IP, which
would extend the ATLAS luminosity detector program at 240~m, is 
currently under investigation by several ATLAS groups. 
Near-beam detectors at $\pm 220$~m from the CMS IP are foreseen as part of the 
TOTEM experiment. The trigger and DAQ systems of TOTEM will be integrated with 
those of CMS such that joint data taking will be possible. The two collaborations
are in the process of defining a joint diffractive and forward physics 
program.\cite{CMSTOTEM}
The TOTEM Silicon detectors are expected to withstand of the order of only 
1~${\rm fb}^{-1}$ of integrated luminosity.  
In order to use detectors at $\pm 220$~m in routine CMS data taking and 
at high luminosities, the TOTEM Silicon detectors will require replacement with
more radiation hard detectors, an option that is currently under investigation.

\section{Status of the FP420 R\&D project}

Design and prototyping of the FP420 detectors and their mechanical and electrical 
support systems are well underway. Engineering studies are well advanced 
to establish the 
necessary modifications to the existing cryogenic LHC elements at 420~m in order 
to accommodate detectors operated at room temperature.
The remaining R\&D work is fully funded till the middle of 2007. 

The FP420 collaboration plans to provide a detailed Technical Design Proposal
to the ATLAS and CMS collaborations in the first half of 2007. 
If ATLAS and/or CMS decide to build the detectors FP420 suggests, Technical Design Reports could be 
provided to the LHCC in 2007.

Installation of the proposed detectors could take place well after the LHC startup
phase, during the first long break in the LHC running schedule, possibly in 
2009/2010.

\section{Conclusions}

Measuring diffractive events at the LHC at luminosities where pile-up is present
requires forward proton tagging capabilities. A radipity gap based selection is
no longer possible. The FP420 R\&D project at CERN aims at providing the 
appropriate means: radiation-hard Silicon detectors located in the cold region
of the LHC at $\pm 420$~m from the ATLAS or CMS IP, complemented with
fast timing detectors to reject fake diffractive events with protons from 
coincidences with pile-up events. Ways of adding detectors at $\pm 220$~m from 
the IP for high-luminosity data taking are under study and discussion in ATLAS, 
CMS and TOTEM.

\section*{Acknowledgments}
The author is supported by the Italian Ministry for Education, University and 
Scientific Research under the program ``Incentivazione alla mobilit\`a di
studiosi stranieri e italiani residenti all'estero''.

\balance

\end{document}